\documentclass[%
aps,
superscriptaddress,
prl,
 amsmath,amssymb,
 reprint,%
author-numerical,%
]{revtex4-2}

\usepackage{xcolor}
\usepackage{graphicx}% Include figure files
\usepackage{dcolumn}% Align table columns on decimal point
\usepackage{bm}% bold math
\usepackage{comment}
\usepackage{ulem}

\begin{document}

\preprint{AIP/123-QED}

%\title{Shaping non-reciprocal caustic spin wave beams from nano-constricted antennas}
\title{Shaping non-reciprocal caustic spin wave beams}

\author{Dinesh Wagle}
\affiliation{Department of Physics and Astronomy, University of Delaware, Newark, Delaware 19716, USA}

\author{Daniel Stoeffler}
 \affiliation{IPCMS - UMR 7504 CNRS Institut de Physique et Chimie des Materiaux de Strasbourg, France}

\author{Loic Temdie}
 \affiliation{IMT Atlantique, Dpt. MO, Lab-STICC - UMR 6285 CNRS, Technopole Brest-Iroise CS83818, 29238 Brest Cedex 03, France}

\author{Mojtaba Taghipour Kaffash}
 \affiliation{Department of Physics and Astronomy, University of Delaware, Newark, DE 19716, USA}
 
 \author{Vincent Castel}
  \affiliation{IMT Atlantique, Dpt. MO, Lab-STICC - UMR 6285 CNRS, Technopole Brest-Iroise CS83818, 29238 Brest Cedex 03, France}

 \author{Hicham Majjad}
 \affiliation{IPCMS - UMR 7504 CNRS Institut de Physique et Chimie des Materiaux de Strasbourg, France}

 \author{Romain Bernard}
 \affiliation{IPCMS - UMR 7504 CNRS Institut de Physique et Chimie des Materiaux de Strasbourg, France}

 \author{Yves Henry}
 \affiliation{IPCMS - UMR 7504 CNRS Institut de Physique et Chimie des Materiaux de Strasbourg, France}

\author{Matthieu Bailleul}
 \affiliation{IPCMS - UMR 7504 CNRS Institut de Physique et Chimie des Materiaux de Strasbourg, France} 
 
\author{M. Benjamin Jungfleisch}
\email{mbj@udel.edu}
 \affiliation{Department of Physics and Astronomy, University of Delaware, Newark, DE 19716, USA}
 
 \author{Vincent Vlaminck}
 \email{vincent.vlaminck@imt-atlantique.fr}
 \affiliation{IMT Atlantique, Dpt. MO, Lab-STICC - UMR 6285 CNRS, Technopole Brest-Iroise CS83818, 29238 Brest Cedex 03, France}

\date{\today}% It is always \today, today,

\begin{abstract}

Caustics are intricate and challenging to control near-field interference patterns that exist in a wide range of physical systems, and which usually exhibit a reciprocal wave propagation. Here, we utilize the highly anisotropic dispersion and inherent non-reciprocity of a magnonic system to shape non-reciprocal emission of caustic like spin wave beams in an extended yttrium iron garnet (YIG) film from a nano-constricted waveguide. We introduce a near-field diffraction model to study spin wave beamforming in homogeneous in-plane magnetized thin films, and reveal the propagation of non-reciprocal spin wave beams directly emitted from the nanoconstriction by spatially resolved microfocused Brillouin light spectroscopy. The experimental results agree well with both micromagnetic simulation, and the near-field diffraction model. 
The proposed method can be readily implemented to study spin wave interference at the sub-µm scale, which is central to the development of wave-based computing applications and magnonic devices.

\end{abstract}

\keywords{Spin wave, Caustics}%Use showkeys class option if keyword
\maketitle

%\section{\label{sec:intro} Introduction}

%Caustics are mathematical concepts describing beamforming 
Caustics is a commonly used term to describe concentration of wave intensity influenced by curvlinear landscape of diffusion or emission. They have been a subject of curiosity in a variety of physical systems, ranging from optics \cite{Kravtsov1993} and dark matter physics \cite{DAVYDOV_2023,White_2009} to condensed matter physics, including phononics \cite{Every1986,Maris1971,Taylor1969}, plasmonics \cite{Shi_2015,Epstein_2014}, electronics \cite{Spector1990, Cheianov2007, Cserti2007}, and magnonics \cite{Bttner2000,Veerakumar2006,Demidov2009,Schneider2010,Kostylev_2011,Sebastian_2013,Gieniusz_2013,Madami2018,Shiota_2020,Muralidhar_2021,Gallardo2021,Wartelle_2023}. While the concept of caustics in optics refers to the concentration of light rays due to reflection or refraction within a heterogeneous system, caustic beams in condensed matter physics are related to the anisotropy of the dispersion relation in a homogeneous system, where the direction of the group velocity and wavevector do not coincide. More specifically, the existence of inflection points in the isofrequency curve %-- also known as the slowness curve in optics -- 
of an anisotropic system leads to a range of wavevectors with group velocity pointing in the same direction. This low spread in group velocity directions around these inflection points results in the generation of a caustic beam.  \\ 
In the last decade, caustics in spin systems have been extensively studied theoretically \cite{Bttner2000,Veerakumar2006,Kim2016,Wartelle_2023}, and experimentally for their advantages with respect to the development of wave-based computing applications such as reservoir computing \cite{Papp2021, Lee2022, Lee2023}, holographic memory \cite{Khitun2013, Khitun2015}, and neuromorphic computing \cite{Papp2021-neural, Christensen2022}. They appear as a promising effect to shape and steer well-defined spin wave beams %\sout{in a narrow} 
in adjustable frequency bands, and with a higher power density than conventional plane waves. The most established technique to excite well-defined caustic beams is to launch spin waves in a narrow ferromagnetic waveguide that enters a semi-infinite plane \cite{Demidov2009,Schneider2010,Heussner2018}, where the junction acts as a pointlike diffracting source. Other methods to create spin wave caustics include the utilization of a collapsing bullet mode \cite{Kostylev_2011}, nonlinear higher harmonic generation from localized edge modes \cite{Sebastian_2013}, or are based on all-optical pointlike sources based on frequency comb rapid demagnetization \cite{Muralidhar_2021}. Recently, the excitation of spin wave caustics has been achieved in extended films from the edges of a straight segment via NV magnetometry \cite{Bertelli_2020}, using a periodic diffraction grating \cite{Makartsou2024}, or in a circular stripline antenna in both the Damon-Eshbach \cite{Wartelle_2023}, and the backward-volume wave geometry \cite{Madami2018}. However, the unique potential of magnonic systems to combine non-reciprocity with nanoscale spin wave beam shaping in an extended thin film remains to be demonstrated.\\
Here, we show the non-reciprocal emission of caustic spin wave beams in an extended yttrium iron garnet (YIG) film from a nano-constricted rf waveguide. We extend the previously developed near-field diffraction model (NFD) of spin waves in out-of-plane magnetized films \cite{Vlaminck2023,Temdie2024} to in-plane magnetized films, which we use to identify suitable antenna designs for shaping spin wave caustic beams. The predictions are then experimentally tested using microfocused Brillouin light scattering (BLS) by mapping the spin wave beam emission. Our experiments reveal a non-reciprocity in the caustic emission that depends on the relative orientation between microwave and biasing magnetic fields. The experimentally obtained spin wave maps agree well with NFD modeling and micromagnetic simulations. Our results highlight the possibility to control the shaping and steering of spin wave caustic beams, which feature narrow adjustable bandwidth, and long-range propagation. %\textcolor{green}{should go in the conclusion instead}\\

\begin{figure}
%\begin{adjustwidth}{-\extralength}{0cm}
\centering
\includegraphics[width=9cm]{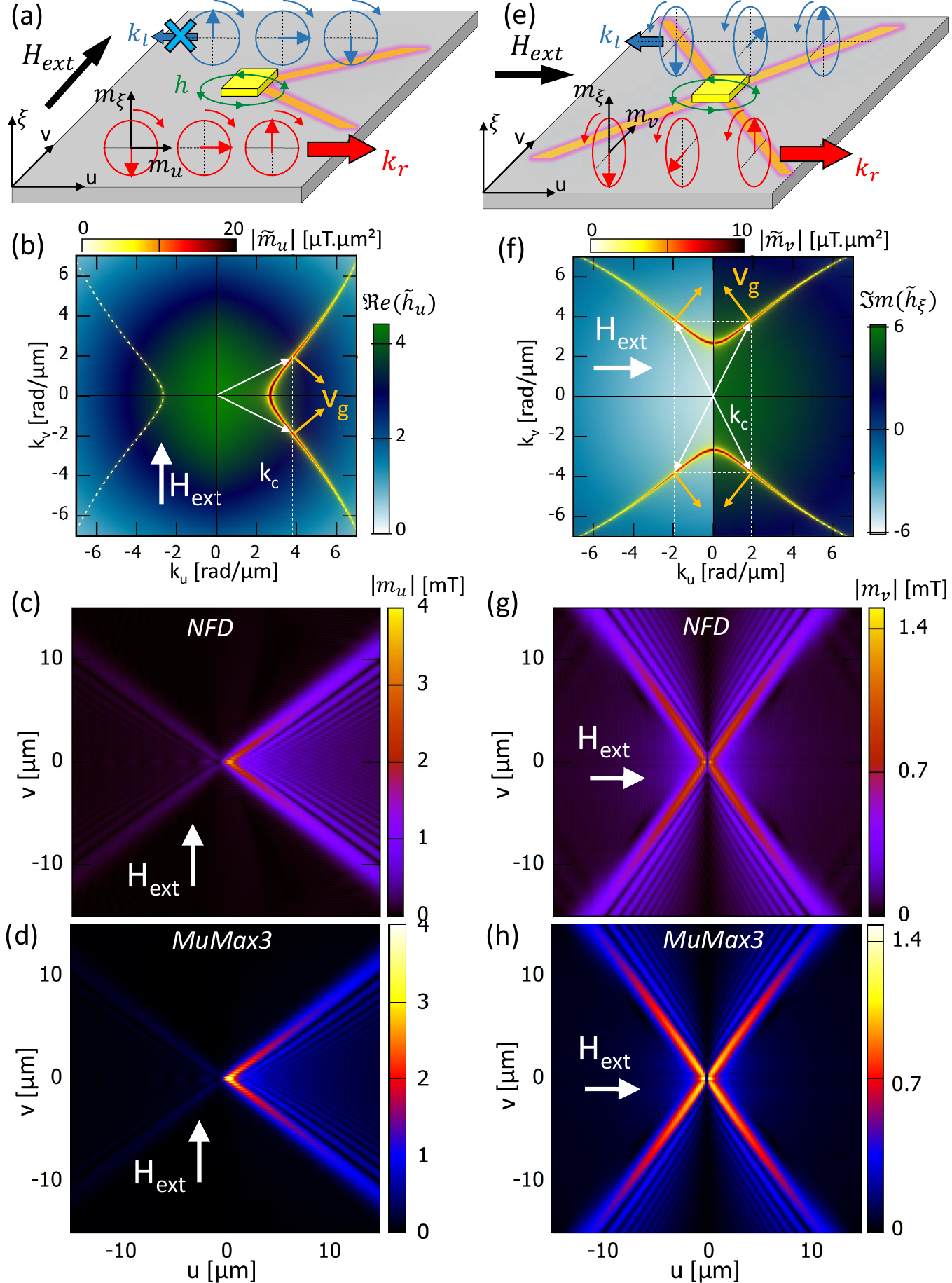}
%\end{adjustwidth}
\caption{(a) Sketch of the spin wave profiles along half a wavelength for oppositely propagating waves in DE configuration, together with the ac field generated by the antenna.(b) Superposition of in-plane Fourier components of the ac magnetization $\lvert\tilde{m}_u\rvert$ (black-red scale bar), and ac field $\mathfrak{Re}(\tilde{h}_u)$ (green-blue scale bar) for a 500\,nm square segment (white dotted line is the isofrequency curve). (c) NFD, and (d) MuMax3 simulations in the DE-like configuration at 7.9\,GHz for a 50\,nm thick YIG film magnetized along $v$ axis under a bias field of 200~mT. (e),(f) Same as (a), (b) for bias field along along $u$-axis, showing out-of-plane ac field $\mathfrak{Im}(\tilde{h}_{\xi})$. (g) MuMax3, and (h) NFD simulations in the BVW-like configuration.
\label{Fig1}}
\end{figure}

%\section{\label{sec:NFD} Near Field diffraction model}

Experimentally, the excitation of coherent spin waves is achieved within a range of wavevector for which the dynamic dipolar interaction usually plays a major role (e.g. $k\leq$100 rad/$\mu$m). For thin films magnetized in-plane, it is specifically this dipolar term that leads to an anisotropic dispersion relation, where the frequency of the spin waves, among other parameters, depends on the relative orientation between magnetization and wavevector \cite{Kalinikos1986}. 
For suitable combinations of field, frequency and wavevector, %\sout{the slowness curve, which is} 
the isofrequency curve in reciprocal space ($k_u,k_v$) possesses an inflection point $\vec{k_c}$, which is referred to as a caustic point $\frac{d^2k_v}{dk_u^2}|_{k=k_c}=0$.
In the vicinity of this point $k_c$ of zero curvature, a finite range of wavevectors have a group velocity pointing in the same direction [$\vec{v_g}=\vec{\nabla}_k(\omega)$], which can lead to the emission of a spin wave caustic beam. The larger the region where the condition $\frac{d^2k_v}{dk_u^2}\approx0$ extends, the easier it will be to excite the caustic beams. Besides, a sufficiently constrained (ideally pointlike) spin wave source is required in order to effectively shape caustic beams.
Figures~\ref{Fig1}(b),~\ref{Fig1}(f) show the isofrequency curves (white dotted lines) at 7.9\,GHz for a 50\,nm-thick YIG film magnetized by a bias field of 200~mT along the $v$- and $u$-axis respectively, superimposed on the Fourier transform of the ac field from a fictitious spin wave antenna consisting of a 500\,nm side square in which a spatially uniform microwave current is assumed to flow along the v axis [Fig.~\ref{Fig1}(a),~\ref{Fig1}(e)].
One can see the necessary condition of having the caustic point located within an effective region of the two-dimensional (2D) emission spectrum in order to generate well-defined beams; see also the Supplemental Material (SM) \cite{SM/SNRCSWB-1} for the comparison with a rectangular segment.\\

In the following, we adapt the NFD approach, which was shown to benchmark spin wave diffraction in out-of-plane magnetized films \cite{Vlaminck2023}, for investigating the beamforming in in-plane configurations. The approach consists of defining the dynamic susceptibility tensor $\bar{\bar{\chi}}_{xy}$ in reciprocal space, which is obtained by inverting the linearized Landau-Lifschitz-Gilbert (LLG) equation \cite{Kalinikos1981,Kalinikos1986} in the 2D frame ($x,y$) transverse to the bias field direction ($z$). For simplicity, we ignore magnetocrystalline anisotropy, and consider unpinned conditions %[$\left( \frac{\partial{\vec{m}}}{\partial{\xi}}\right)_{\pm t/2}=\vec{0}$] 
at both top and bottom surfaces. We also restrict ourselves to the fundamental mode ($n=0$), for which the spin wave profile is uniform across the thickness, and assume no coupling with higher-order modes. This last assumption is all the more relevant the thinner the film is, for which higher-order modes are fairly decoupled. 
\begin{figure*}
%\begin{adjustwidth}{-\extralength}{0cm}
\centering
\includegraphics[width=18cm]{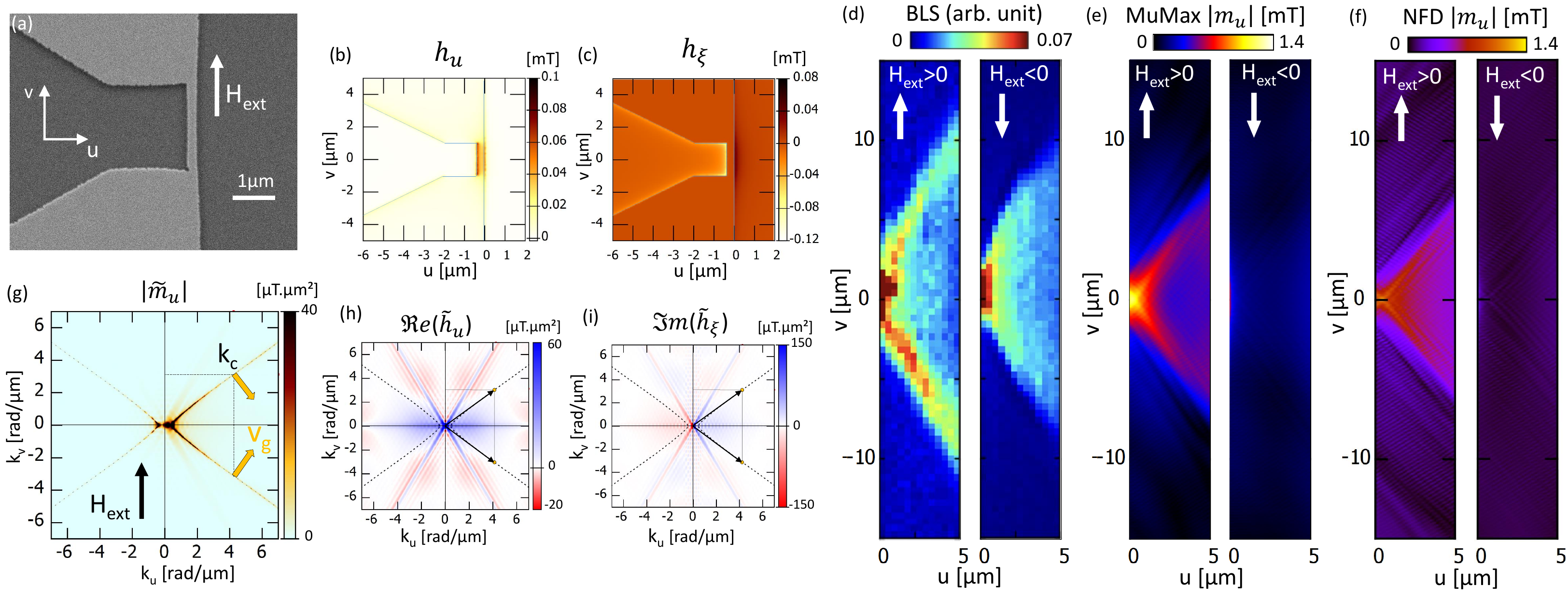}
%\end{adjustwidth}
\caption{(a) Scanning microscope image of a 400\,nm-wide and 2\,$\mu$m-long constricted stripline. (b) 2D-field distribution of the in-plane $h_u$, and (c) out-of-plane $h_{\xi}$ components of the excitation field obtained via {\scriptsize COMSOL MULTIPHYSICS}.  (d) BLS measurement at 7.5\,GHz, and a bias field $\mu_0 H_\mathrm{ext}$=$\pm$188\,mT applied in the $v$-direction. (e) Corresponding MuMax3 simulations, and (f) NFD simulations for both bias field polarities.  (g) In-plane Fourier components of the dynamic magnetization $\lvert\tilde{m}_u\rvert$. (h) Real part of the in-plane Fourier component of the ac field $\mathfrak{Re}(h_u)$. (i) Imaginary part of the out-of-plane Fourier component of the ac field $\mathfrak{Im}(h_\xi)$ (dotted line is the isofrequency).
\label{Fig2}}
\end{figure*} 

The diffraction pattern from an arbitrarily shaped coplanar waveguide (CPW) can then be obtained by computing the inverse Fourier transform of the matrix product of the susceptibility tensor with the Fourier transform of the excitation field: 
\begin{equation}
\vec{m}(u,v,t)=e^{i\omega t}\iint_{-\infty}^{+\infty}dk_udk_v \,\,\bar{\bar{\chi}}\,\, \vec{h}(k_u,k_v)\,  e^{-i(k_uu+k_vv)} ,
\label{eq1}
\end{equation}
where the susceptibility tensor $\bar{\bar{\chi}}$, and the excitation field $\vec{h}(k_u,k_v)$ have been expressed in the laboratory frame ($u,v$), which is related to the shape of the spin wave antenna \cite{SM/SNRCSWB-1}. $\vec{h}(k_u,k_v)$ is a two-dimensional vector, made of the out-of-plane component $h_{\xi}$, and the in-plane component $h_{in}$ transverse to the bias field direction ($z$). 
Thus, we obtain a mapping in the steady state of the dynamic magnetization vector $\vec{m}(u,v,t)=(m_{\xi},m_{in})$.\\

\begin{figure}
%\begin{adjustwidth}{-\extralength}{0cm}
\centering
\includegraphics[width=8cm]{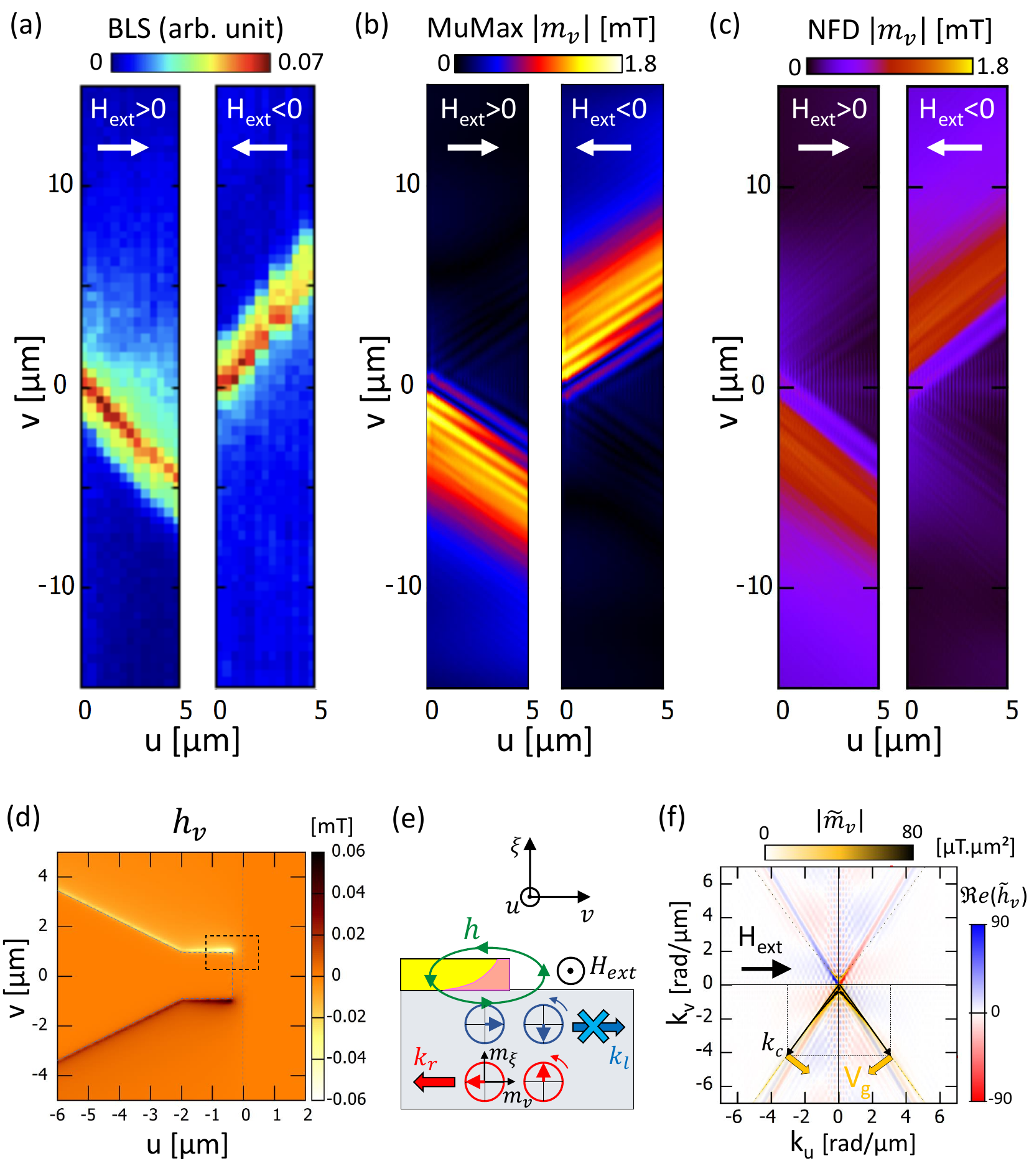}
%\end{adjustwidth}
\caption{(a) BLS measurement at 7.5\,GHz, and a +188\,mT (left) and -188\,mT (right) bias field applied along $u$. (b) Corresponding MuMax3, and (c) NFD simulations for both bias field polarities. (d) In-plane component $h_v$ of the microwave field obtained with {\scriptsize COMSOL MULTIPHYSICS}. (e) Sketch of the chiral coupling occurring at the edges of the constriction in the BVW-like configuration. (f) Superposition of the in-plane Fourier components of ac magnetization $\lvert\tilde{m}_v\rvert$ (black-yellow color scale), and ac field $\lvert h_v\rvert$ (red-blue color scale) indicating the effective caustic locations. 
\label{Fig3}}
\end{figure}

Figure~\ref{Fig1} shows the spin wave diffraction patterns obtained for the 500\,nm square on top of a 50\,nm-thick YIG film with a Gilbert damping $\alpha$=2\,$\times10^{-4}$. %\sout{[see SM for the corresponding results of the 5\,$\mu$m segment]}. 
In these simulations, we used a simple excitation profile derived from the expressions of the Oersted field of a 80\,nm-thick straight rectangular conductor carrying uniform current density along $v$, for which we adjusted the current value to have maximum field of 0.1\,mT. We simulate two limiting cases: (i) when the bias field is perpendicular to the in-plane component of the excitation field [Fig.~\ref{Fig1}(c), Damon-Eshbach (DE) like configuration], and (ii) when the bias field is parallel to the in-plane component of the excitation [(Fig.~\ref{Fig1}(g), backward volume wave (BVW) like configuration]. Alongside the NFD mapping, we also show in Figs.~\ref{Fig1}(d),~\ref{Fig1}(h) the corresponding micromagnetic simulations using MuMax3 \cite{Vansteenkiste_2014}, and taking into account the same set of parameters and excitation geometry \cite{SM/SNRCSWB-1}. 
One can clearly appreciate the excellent quantitative agreement between both methods, 
and hence, these simple tests validate the NFD approach for sufficiently thin in-plane magnetized films. 
Moreover, the comparison between a 5\,$\mu$m-long rectangular segment \cite{SM/SNRCSWB-1}, and the 500\,nm square evidence how a pointlike source is essential for properly shaping a single caustic beam.\\ 
Furthermore, a pronounced non-reciprocity is observed in the DE-like configuration, for which a drastically reduced intensity is found on the left-hand side with respect to the bias field direction [Figs.~\ref{Fig1}(c),~\ref{Fig1}(d)]. As is sketched in Fig.~\ref{Fig1}(a), a chiral coupling occurs, in which the microwave field distribution only matches the phase profile of spin waves for a single-sided propagation \cite{Yu2021,Devolder2023,Temdie2023,Temdie2023_MDPI,SM/SNRCSWB-1}. In essence, the chirality comes from the interplay of the even in-plane, and the odd out-of-plane components of the ac field with respect to center of the stripline. 
As the magnetic response results from the product in reciprocal space between the antisymmetric susceptibility tensor and the excitation field, the wavevectors for which the in-plane and the out-of-plane component of the ac field have the same sign will be more effectively excited \cite{SM/SNRCSWB-1}. We superimposed the in-plane Fourier component of the dynamic magnetization (black-red-yellow color scale) on the in-plane $\mathfrak{Re}(\tilde{h}_u)$, and the out-of-plane $\mathfrak{Im}(\tilde{h}_{\xi})$ components respectively in Figs.~\ref{Fig1}(b),(f). As expected, only the two caustic points on the right-hand side of the bias field ($k_u>0$) are effectively excited in the DE-like configuration. In contrast, the caustic beams appear fully symmetrical on either side of the bias field when it is parallel to the in-plane ac field [Figs.~\ref{Fig1}(g),~\ref{Fig1}(h)], as only the out-of-plane component of the ac field couples to the spin waves, and therefore the matching of the spin waves phase profiles is permitted for both propagation directions as sketched in Figs.~\ref{Fig1}(e), and all four caustic points are symmetrically excited in the BVW-like configuration. In summary, chiral spin waves excitation favors directions to the right-hand side of the bias field direction, accordingly with the Damon-Eshbach product rule $\vec{k}_{\perp}/k_{\perp} = \vec{n}_0 \times \vec{M}/M_s$ \cite{Gurevich1996-qv}, where $n_0$ is the internal normal to the film top (bottom) surface if the antenna sits on top (bottom) of the film. \\
To confirm our theoretical modeling results, we experimentally investigate the generation of caustic beams from a nano-constricted stripline patterned on top of an extended 200-nm thick YIG film by microfocused BLS spectroscopy \cite{SM/SNRCSWB-1}. We fabricated two sets of antenna to study the two different field orientations: (i) DE-like configuration (Fig.~\ref{Fig2}), and (ii) BVW-like configuration (Fig.~\ref{Fig3}). Figure~\ref{Fig2}(a) shows an exemplary scanning electron image of a $L$=2\,µm-long and $w$=400\,nm-wide constriction patterned on the top of the YIG film, using electron beam lithography followed by ebeam evaporation of 6\,nm-Ti/80\,nm-Au.

%\subsection{\label{H_perp_h} Damon-Eshbach-like configuration } 
\textbf{(i) Damon-Eshbach-like configuration.}
We first present the BLS measurement for the geometry where the preponderant in-plane component of the excitation field $h_u$ at the constriction is perpendicular to the bias field. Figure~\ref{Fig2}(d) shows the 2D BLS maps obtained with continuous microwave power of $-3$\,dBm at 7.5\,GHz, and an applied field of $\mu_0 H_\mathrm{ext}$=$\pm$188\,mT. Only measurement on the right-hand side of the constriction were achievable as the BLS laser beam cannot penetrate through the gold antenna covering the left handside. For the positive polarity, e.g., field pointing upward [left panel of Fig.~\ref{Fig2}(d)], we observe two well-defined beams propagating symmetrically away from the constriction at about 58$^{\circ}$ with respect to the $x$-axis. Conversely, for the negative polarity, e.g., field pointing downward [right panel of Fig.~\ref{Fig2}(d)], the beams are much less defined, and the signal rapidly decays as we move away from the constriction. This effect appears more pronounced in the corresponding MuMax3 and NFD simulations, shown respectively in Fig.~\ref{Fig2}(e) and Fig.~\ref{Fig2}(f), for which we used the microwave field distribution obtained with {\scriptsize COMSOL MULTIPHYSICS} as input, and which are shown in Fig.~\ref{Fig2}(b),~\ref{Fig2}(c) \cite{SM/SNRCSWB-1}.\\
This field-orientation non-reciprocity is due to the chiral coupling explained above. 
Figure~\ref{Fig2}(g) displays the in-plane Fourier component of the dynamic magnetization $\lvert\tilde{m}_u\rvert$, which indicates a much larger coupling strength for the wavevector pointing to the right-hand side of the bias field. In fact, the difference of parity between the in-plane and the out-of-plane components of the ac field shown respectively in Fig.~\ref{Fig2}(h) and Fig.~\ref{Fig2}(i) confirms a more effective contribution for $k_u>0$, where both components have the same sign.
Despite some minor differences in chiral coupling between measurements and simulations, which could be due to the difficulty of defining the microwave field distribution of the real device, a reasonable agreement between BLS measurements and simulations is found, demonstrating the control of caustic spin wave beam excitation at the nanoscale directly from an antenna. \\

%\subsection{\label{H_para_h} Backward-volume-wave-like configuration } 
\textbf{(ii) Backward-volume-wave-like configuration.}
In this configuration, the bias field is parallel to the in-plane component $h_u$ of the microwave magnetic field at the constriction. Figure~\ref{Fig3}(a) shows 2D BLS maps obtained using identical experimental conditions as in the DE-like configuration, e.g., an excitation frequency of 7.5\,GHz, and an external field of $\mu_0 H_\mathrm{ext}$=$\pm$188\,mT. As is evident from the figure, a drastically different behavior is observed in the BVW-like configuration. Strikingly, the BLS measurements show a single sharply defined caustic beam emitted from the constriction, which exhibits a non-reciprocal behavior upon bias magnetic field reversal. If the magnetic field is positive along the $u$ axis [$H_\mathrm{ext}>0$, left panel of Fig.~\ref{Fig3}(a)], a single spin wave beam propagating in the negative $v$ direction (downward) is detected. On the other hand, when $H_\mathrm{ext}<0$ [right panel of Fig.~\ref{Fig3}(a)], a single upward propagating spin wave beam is detected. Each beam is comparable in intensity, and both have similar caustic angles $\approx\pm$45$^{\circ}$. We show the corresponding MuMax3 and NFD simulations in Fig.~\ref{Fig3}(b) and Fig.~\ref{Fig3}(c) for which we also used the microwave field distribution obtained via {\scriptsize COMSOL}. Both simulations exhibit similar field-dependent single beam emission (see SM \cite{SM/SNRCSWB-1} for full-scale simulations). \\
We again attribute this field-dependent non-reciprocity in the BVW-like configuration to a chiral coupling, but involving this time the interplay between the out-of-plane component $h_{\xi}$ and the in-plane component $h_v$ of the microwave field. As shown in Fig.~\ref{Fig3}(d), the {\scriptsize COMSOL} simulations reveal that a significant microwave power localized around the edges of the coplanar waveguide, which creates a thin line of in-plane microwave field $h_v$ perpendicular to the bias field close to the constriction. We sketched in Fig.~\ref{Fig3}(e) the essence of the chiral coupling in this configuration (bias field in the $u$ direction), where the transverse propagation of spin waves in the $v$ direction is only allowed on the right-hand side of the bias field. We also show in Fig.~\ref{Fig3}(f) the overlaying of the in-plane Fourier components of ac magnetization $\lvert\tilde{m}_v\rvert$ with the FFT of the $h_v$ component of the microwave field confirming an efficient excitation only for the two caustic points located on the right-hand side of $\vec{H_\mathrm{ext}}$.
Besides the satisfying agreement between measurement and simulations, one notices that the measured beam appears slightly thinner than in the simulations, which is likely due to the difficulty of reproducing the microwave field distribution of the real device.\\

\begin{figure}
%\begin{adjustwidth}{-\extralength}{0cm}
\centering
\includegraphics[width=8cm]{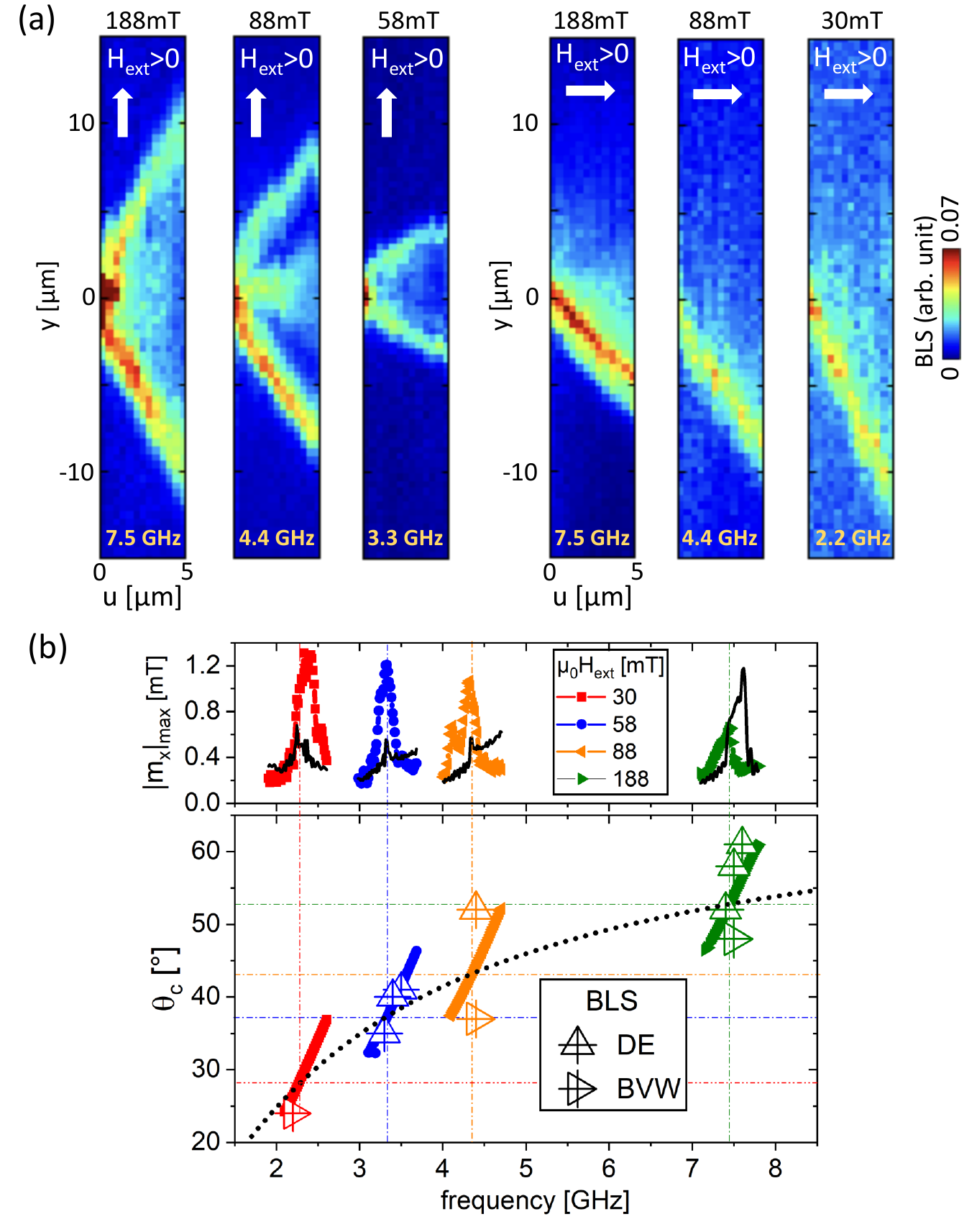}
%\end{adjustwidth}
\caption{(a) BLS measurement showing the angular dependence of the field and frequency. (b) Summary of caustic existence conditions. (Top) Frequency dependence of amplitude $|m_x|_\mathrm{max}$, which corresponds to the maximum amplitude of $|m_x|$ at $u$=10\,$\mu$m in the NFD simulations (black lines are for BVW-configuration, colored lines for DE-configuration). (Bottom) Comparison between measured caustic angles and theoretical values $\theta_c=\pi/2-(\widehat{\vec{v_g},\vec{H}_{ext}})$ extracted from the isofrequency curve at the corresponding applied fields. (black dashed line corresponds to FMR condition in field and frequency). 
\label{Fig4}}
\end{figure}

%\subsubsection{\label{Caustic_prop} Beam steering} 
%\textbf{Beam steering}
In the following, we analyze the angular dependence of the beam formation as a function of the bias field and excitation frequency, as well as the frequency dependence of the beam intensity at a given field. Figure~\ref{Fig4}(a) shows 2D BLS maps at several bias fields (188, 88, 58, 30)\,mT, respectively at (7.5, 4.4, 3.3, 2.2)\,GHz for both configurations. We observe that the beams can be gradually steered toward the bias field direction as the frequency is increased. In the DE-like configuration, the angle of the caustic beam with respect to the $u$ axis increases with frequency, while in the BVW-like configuration, the angle decreases with increasing frequency. Figure~\ref{Fig4}(b) compares the experimentally obtained angles with the theoretically expected caustic angles, derived numerically from $f(k_u,k_v)$ as the angle between the group velocity and the direction perpendicular to the bias field \cite{SM/SNRCSWB-1}. We find a reasonable agreement within the error bars for the DE-like configuration, while the measured angles are systematically lower than the theoretically expected ones for the BVW-like configuration. \\
%We suspect that the actual microwave field distribution has a stronger effect on the short-range shape of the beam in this configuration. 
Additionally, we performed 50 NFD simulations for each bias field value over a 700~MHz frequency span to plot the frequency dependence of the caustic-beam amplitude, as presented in the top panel of Fig.~\ref{Fig4}(b). Strikingly, we find that the caustic-beam amplitude peaked right around the field-frequency values corresponding to the ferromagnetic resonance (FMR) condition. This observation is consistent with the fact that the zero curvature ($d^2k_v/dk_u^2\sim~0$) of the isofrequency curve extends over a broader range of wavevectors at around the FMR frequency for a given field. Furthermore, one notices that the peak amplitude of the caustic beams decreases with frequencies in the DE-configuration whereas it increases in the BVW-configuration [compare black and colored lines in the top panel of Fig. 4(b)].

%\section{Conclusion}

In conclusion, we demonstrated theoretically and experimentally the possibilty to shape non-reciprocal caustic beams in homogeneous ferromagnetic thin film directly from a nano-constricted stripline. First, we introduced a model that efficiently and accurately predicts spin wave diffraction in thin film magnetized in-plane, and which can readily predict the shaping of caustic spin wave beams from arbitrary shaped antenna. Then, using microfocused BLS, we confirmed the predictions of the model, revealing spin wave caustic chiral emission from nanometric constrictions. We studied two measurement configurations, one for which the in-plane microwave field was essentially perpendicular to the magnetization (DE) throughout the constriction, and the other one for which they were parallel to each other (BVW). In the DE configuration, we observed a chiral excitation of two spin wave beams for a given field polarity. In the BVW configuration, we observed a single non-reciprocal upward- or downward-going beam that can be controlled at ease by changing the polarity of the bias field, and with a steerability precisely controlled by tuning the bias field magnitude and the excitation frequency. Our work provides deep insight into tailoring spin wave caustics using nano-sized waveguides, which we expect will accelerate the development of interference-based magnonic logic and computing devices.\\

\begin{acknowledgments}
This work has benefited from a government grant operated by the French National Research Agency as part of the France 2030 program, Reference No. ANR-22-EXSP-0004 (SWING), as well as the ANR project \textit{MagFunc}/ANR-20-CE91-0005, and the Transatlantic Research Partnership, a program of FACE Foundation and the French Embassy. We also acknowledge financial support by the Interdisciplinary Thematic Institute QMat, as part of the ITI 2021-2028 Program of the University of Strasbourg, CNRS and Inserm, IdEx Unistra (ANR 10 IDEX 0002), SFRI STRAT’US Project (ANR 20 SFRI 0012), and ANR-17-EURE-0024 under the framework of the French Investments for the Future Program, as well as the High Performance Computing Center of the University of Strasbourg for supporting this work by providing scientific support and access to computing resources. Part of the computing resources were funded by the Equipex Equip@Meso project (Programme Investissements d'Avenir) and the CPER Alsacalcul/Big Data. Research at the University of Delaware was supported by the U.S. Department of Energy, Office of Basic Energy Sciences, Division of Materials Sciences and Engineering under Award No. DE-SC0020308. The authors acknowledge the use of facilities and instrumentation supported by NSF through the University of Delaware Materials Research Science and Engineering Center, DMR-2011824. We thank Anish Rai for characterizing the YIG films by FMR.
\end{acknowledgments}

\nocite{*}
\bibliography{CausticNFD}% Produces the bibliography via BibTeX.

\end{document}